\documentclass{ws-ijbc}
\usepackage{ws-rotating}     
\usepackage{url}
\usepackage{subfigure}

\begin{document}

\catchline{}{}{}{}{} 

\markboth{Genaro J. Mart{\'i}nez, Andrew Adamatzky, Ramon Alonso-Sanz}{Complex dynamics of elementary cellular automata emerging from chaotic rules}

\title{Complex dynamics of elementary cellular automata emerging from chaotic rules}

\author{Genaro J. Mart{\'i}nez}
\address{Instituto de Ciencias Nucleares, and Centro de Ciencias de la Complejidad \\ Universidad Nacional Aut\'onoma de M\'exico, M\'exico. \\  Unconventional Computing Center, Bristol Institute of Technology, University of the West of England, United Kingdom.
\\ genaro.martinez@uwe.ac.uk}

\author{Andrew Adamatzky}
\address{Unconventional Computing Center, Bristol Institute of Technology, University of the West of England, Bristol, United Kingdom. \\ andrew.adamatzky@uwe.ac.uk}

\author{Ramon Alonso-Sanz}
\address{ETSI Agr\'onomos, Polytechnic University of Madrid, Madrid, Spain. \\ Unconventional Computing Center, Bristol Institute of Technology, University of the West of England, United Kingdom. \\ ramon.alonso@upm.es}

\maketitle

\begin{history}
\received{(to be inserted by publisher)}
\end{history}

\begin{abstract}
We show techniques of analyzing complex dynamics of cellular automata (CA) with chaotic behaviour. CA are well known computational substrates for studying emergent collective behaviour, complexity, randomness and interaction between order and chaotic systems. A number of attempts have been made to classify CA functions on their space-time dynamics and to predict behaviour of any given function. Examples include mechanical computation, $\lambda$ and $Z$-parameters, mean field theory, differential equations and number conserving features. We aim to classify CA based on their behaviour when they act in a historical mode, i.e. as CA with {\it memory}. We demonstrate that cell-state transition rules enriched with memory quickly transform a chaotic system converging to a complex global behaviour from almost any initial condition. Thus just in few steps we can select chaotic rules without exhaustive computational experiments or recurring to additional parameters. We provide analysis of well-known chaotic functions in one-dimensional CA, and decompose dynamics of the automata using majority memory exploring glider dynamics and reactions.
\end{abstract}



\section{Introduction}
\noindent In this paper we consider a simple tool to extract complex systems from a family of chaotic discrete dynamical system. We will employ a technique --- {\it memory} based rule analysis \cite{AM03}, \cite{Alo08}, \cite{Alo09}, \cite{Alo09a}, of using past history of  a system to construct its present state and to manipulate its future.

We focus on one-dimensional CA. CA are well known computational substrates for studying emergent collective behaviour, complexity, randomness and interaction between order and chaos. A number of efforts have been made to classify CA functions on their space-time dynamics and to predict behaviour of any given function. Examples include mechanical computation, $\lambda$ and $Z$-parameters, mean field theory, differential equations and number conserving features. We aim to base CA classification on their behaviour in a historical mode, i.e. as {\it CA with memory} \cite{Alo08}.

We study elementary CA (ECA) where each function evaluates a central cell with their two neighbourhoods (left and right) and every cell takes a value of its binary alphabet. ECA are introduced and extensively studied by \cite{Wolf84}, \cite{Wolf02}. In ECA there is a set of functions determining global chaotic behaviour where global configurations are chaotic, many configurations have  many ancestors, and attractors are dense \cite{WL92}.

ECA is a one-dimensional array of finite automata, each automaton takes two states and updates its state in discrete time depending on its own state and states of its two closest neighbours, all cells update their state synchronously. A general classification of ECA was introduced in \cite{Wolf94}, as follows:

\begin{itemlist}
\item {class I.} CA evolving to a homogeneous state.
\item {class II.} CA evolving periodically.
\item {class III.} CA evolving chaotically.
\item {class IV.} Include all previous cases, as well known as class of {\it complex rules}.
\end{itemlist}

In this classification class IV is of particular interest because the rules of the class  exhibit non-trivial behaviour with rich diversity of patterns emerging and non-trivial interactions between travelling localizations, or gliders, e.g. ECA Rule 54 \cite{MAM06}.

In present paper we aim to  transform a chaotic evolution rule to a complex system by using memory

\begin{center}
{\small chaotic ECA $\xrightarrow{\mbox{memory}}$ complex ECA}
\end{center}
 
\noindent and derive a new classes of CA functions with historic evolution.

We believe that by employing historic evolution we are able to explore hidden properties of chaotic systems, and select chaotic rules with homogeneous dynamics.

\section{Basic notation}

\subsection{One-dimensional cellular automata}

One-dimensional CA is represented by an array of {\it cells} $x_i$ where $i \in Z$ (integer set) and each $x$ takes a value from a finite alphabet $\Sigma$. Thus, a sequence of cells \{$x_i$\} of finite length $n$ represents a string or {\it global configuration} $c$ on $\Sigma$. This way, the set of finite configurations will be represented as $\Sigma^n$. An evolution is represented by a sequence of configurations $\{c_i\}$ given by the mapping $\Phi:\Sigma^n \rightarrow \Sigma^n$; thus their global relation is following

\begin{equation}
\Phi(c^t) \rightarrow c^{t+1}
\label{globalFunction}
\end{equation}

\noindent where $t$ is time steps and every global state of $c$ is defined by a sequence of cell states. Also the cell states in configuration $c^t$ are updated at the next configuration $c^{t+1}$ simultaneously by a local function $\varphi$ as follows

\begin{equation}
\varphi(x_{i-r}^t, \ldots, x_{i}^t, \ldots, x_{i+r}^t) \rightarrow x_i^{t+1}.
\label{localFunction}
\end{equation}

Following \cite{Wolf84}, \cite{Wolf02} one can represents any CA with two parameters $(k,r)$. Where $k = |\Sigma|$ is a number of states, and $r$ is a radius of neighbourhood. Thus ECA are defined by parameters $(2,1)$. There are $\Sigma^n$ different neighbourhoods (where $n=2r+1$) and $k^{k^n}$ different evolution rules.

In computer experiments we are using automata with periodic boundary conditions.

\subsection{Cellular automata with memory}
Conventional cellular automata are ahistoric (memoryless). A new state of a cell depends on the neighbourhood configuration solely at the preceding time step of $\varphi$ (see Eq.~\ref{localFunction}).

CA with {\it memory} extends standard framework of CA by allowing every cell $x_i$ to remember some period of its previous evolution \cite{Alo08}.

Thus to implement a memory we design a memory function $\phi$, as follows:

\begin{equation}
\phi (x^{t-\tau}_{i}, \ldots, x^{t-1}_{i}, x^{t}_{i}) \rightarrow s_{i}
\end{equation}

\noindent such that $\tau < t$ determines the degree of memory backwards and each cell $s_{i} \in \Sigma$ is a state function of the series of states of the cell $x_i$ with memory up to time-step. To execute the evolution we apply the original rule as follows:

$$
\varphi(\ldots, s^{t}_{i-1}, s^{t}_{i}, s^{t}_{i+1}, \ldots) \rightarrow x^{t+1}_i.
$$

In CA with memory,  while the mapping $\varphi$ remains unaltered, historic memory of all past iterations is retained by featuring each cell as a summary of its past states from $\phi$. Therefore cells {\it canalize} memory to the map $\varphi$.

\begin{figure}[th]
\centerline{\includegraphics[width=5in]{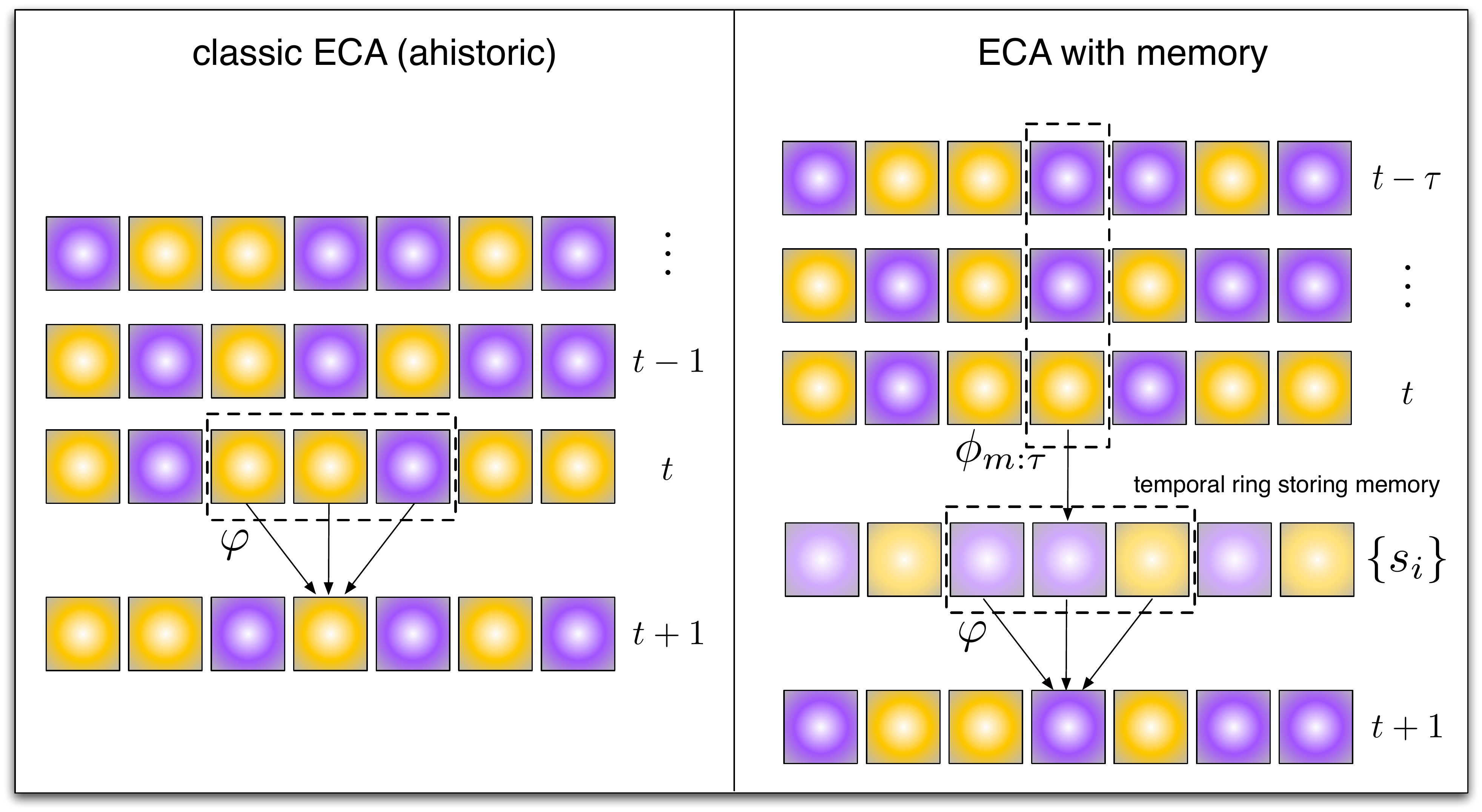}}
\caption{Cellular automata with memory in cells.}
\label{memEvol}
\end{figure}

For example, let us consider memory function $\phi$ as a {\it majority memory}:

\begin{equation}
\phi_{maj} \rightarrow s_{i}
\label{eq-majmem}
\end{equation}

\noindent where in case of a tie given by $\Sigma_1 = \Sigma_0$ in $\phi$, we shall take the last value $x_i$. So $\phi_{maj}$ represents the classic majority function for three variables \cite{Mins67}, hence we have:

\begin{equation*}
\phi_{maj} : (x_1 \wedge x_2) \vee (x_2 \wedge x_3) \vee (x_3 \wedge x_1) \rightarrow x
\end{equation*}

\noindent on cells $(x^{t-\tau}_{i}, \ldots, x^{t-1}_{i}, x^{t}_{i})$ and define a temporal ring before calculating the next global configuration $c$. The representation of a ECA with memory \cite{MAA10} is given as follows:

\begin{equation}
\phi_{CARm:\tau}
\end{equation}

\noindent where $CAR$ is a decimal notation of a particular ECA rule and $m$ the kind of memory given with a specific value of $\tau$. Thus the majority memory ($maj$) working in ECA Rule 86 checking tree cells ($\tau=3$) of history is simply denoted as: $\phi_{R86maj:3}$. Figure~\ref{memEvol} depicts in detail the memory effect working on ECA.

Note that memory is a simple function but its global behaviour $\Phi$  can be predicted from its local function $\phi$ and $\varphi$.

\section{Classes of ECA by polynomials}

\subsection{Mean filed approximation}

Mean field theory is a proven technique for discovering general statistical properties of CA without analyzing evolution spaces of individual rules \cite{Mc09}.

The method assumes that elements of the set of states $\Sigma$ are independent, uncorrelated between each other in the rule's evolution space $\varphi$. Therefore we can study probabilities of states in neighbourhood in terms of probability of a single state (the state in which the neighbourhood evolves), thus probability of a neighbourhood is the product of the probabilities of each cell in the neighbourhood.

In this way, it was proposed to explain Wolfram's classes by a mixture of probability theory and de Bruijn diagrams in \cite{Mc90}, resulting in a classification based on mean field theory curve:

\begin{itemlist}
\item class I: monotonic, entirely on one side of diagonal;
\item class II: horizontal tangency, never reaches diagonal;
\item class IV: horizontal plus diagonal tangency, no crossing;
\item class III: no tangencies, curve crosses diagonal.
\end{itemlist}

Thus for one dimension all cell neighbourhoods must be considered as:

\begin{equation}
p_{t+1}=\sum_{j=0}^{k^{2r+1}-1}\varphi_{j}(X)p_{t}^{v}(1-p_{t})^{n-v}
\label{MFp1D}
\end{equation}

\noindent such that $j$  is a number of relations from neighbourhoods and $X \in \Sigma$ represent of cells $x_{i-r},\ldots,x_{i},\ldots,x_{i+r}$. Thus $n$ represents the number of cells in neighborhood, $v$ indicates how often state one occurs in the neighborhood, $n-v$ shows how often state zero occurs in the neighborhood, $p_{t}$ is a probability of cell being in state one, $q_{t}$ is a probability of cell being in state zero (such that $q=1-p$).

\section{Complex dynamics emerging from chaotic ECA}

\subsection{Chaotic ECA}
Let us consider two cases of classic ECA with chaotic behaviour to demonstrate our results: the evolution rules 86 and 101.

\begin{figure}[th]
\begin{center}
\subfigure[]{\scalebox{0.65}{\includegraphics{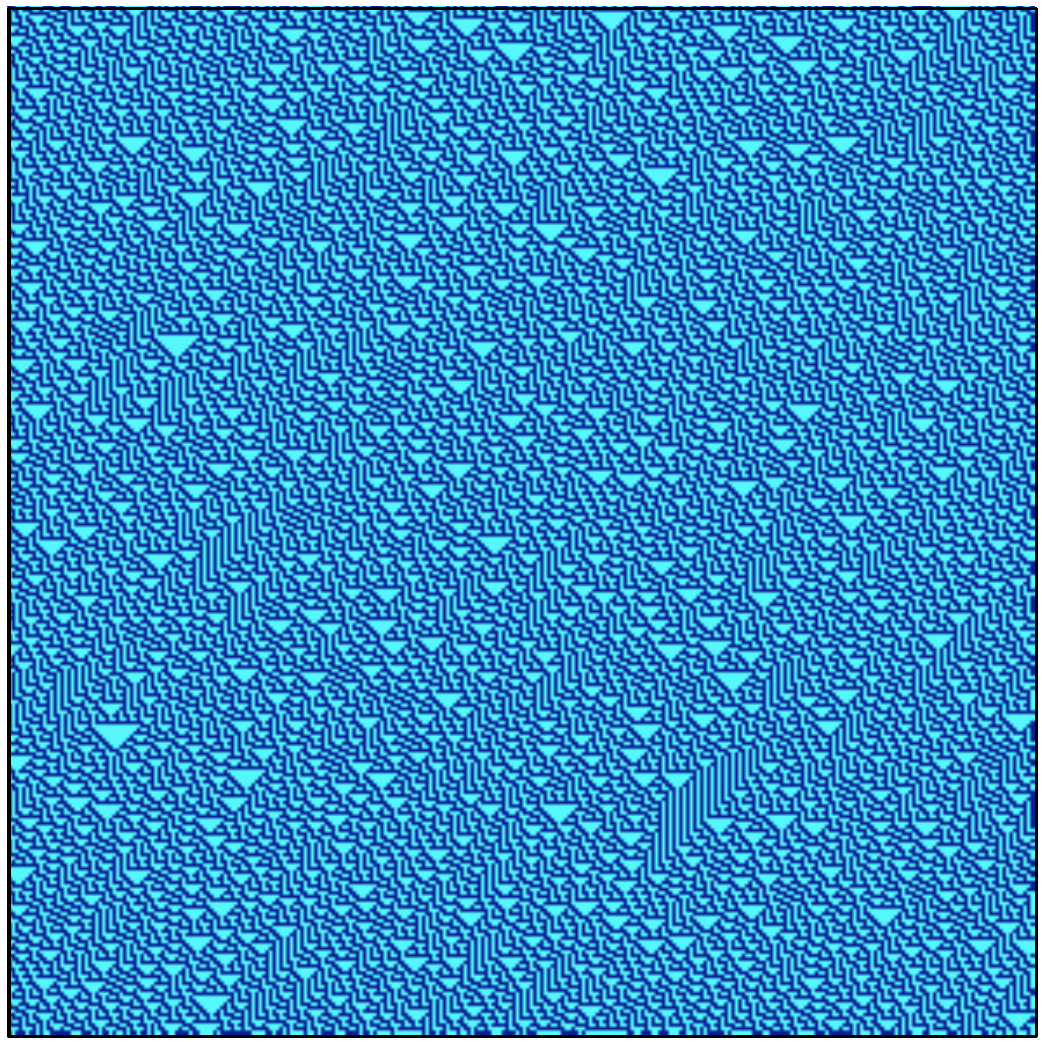}}} \hspace{1.5cm}
\subfigure[]{\scalebox{0.65}{\includegraphics{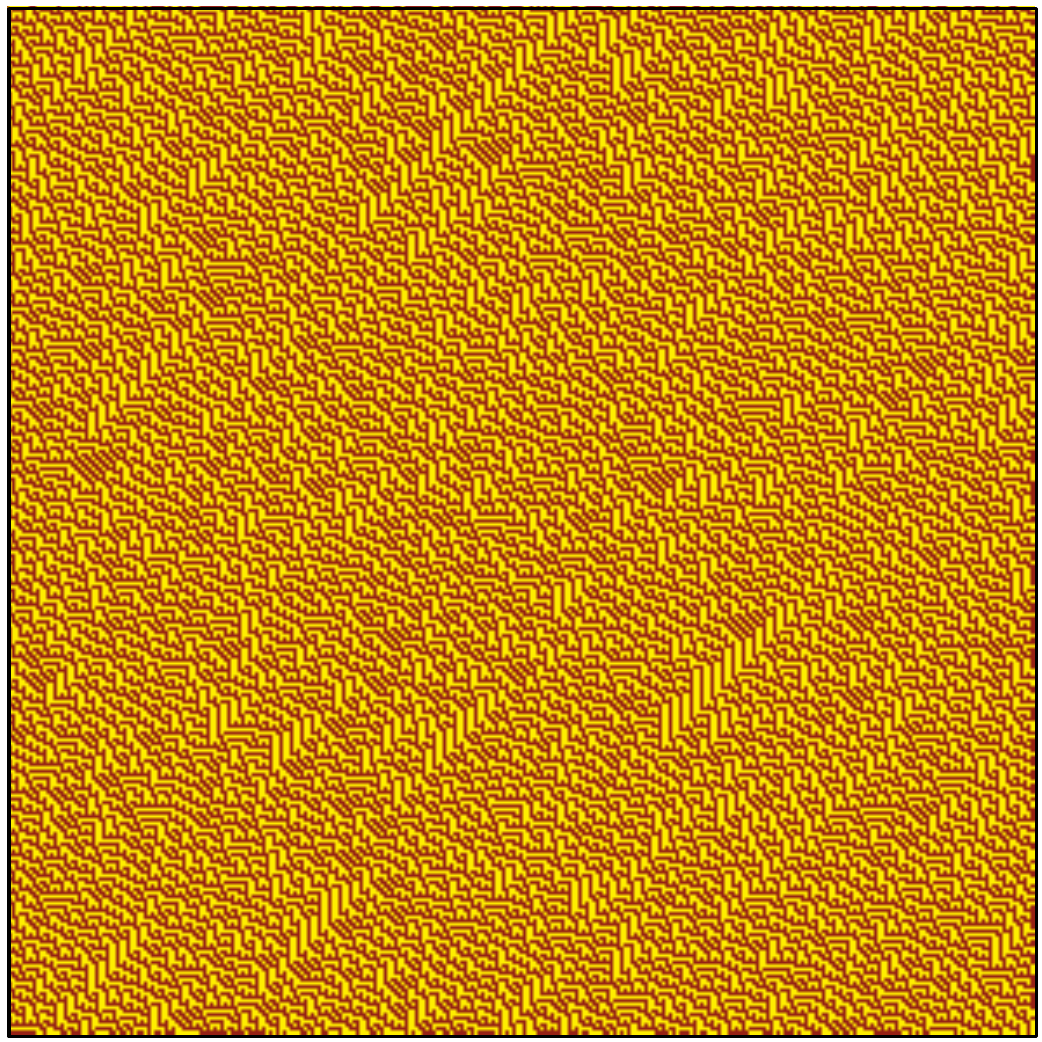}}}
\end{center}
\caption{Chaotic global behaviour in ECA evolution rules (a) $\varphi_{R86}$ and (b) $\varphi_{R101}$ evolving over an array of 295 cells in 295 generations. Both evolutions start in configurations with the same random initial density of 50\%.}
\label{chaosECA}
\end{figure}

We need to provide their mean filed approximation to verify that both function have a chaotic global behaviour before selecting the memory.

The local rule $\varphi$ corresponding to rule 86 is following:

\[
\varphi_{R86} = \left\{
	\begin{array}{lcl}
		1 & \mbox{if} & 110, 100, 010, 001 \\
		0 & \mbox{if} & 111, 101, 011, 000
	\end{array} \right. .
\]

Initially $\varphi_{R86}$ has produces states zero and one with the same probability. There is an equilibrium of states in $\Phi$. On the other hand, $\varphi_{R86}$ determines a surjective correspondence and therefore every configuration has at least one ancestor and no Garden of Eden configurations [Amoroso \& Cooper, 1970]. Of course this rule is the reflection of well-known ECA rule 30 \cite{WL92}.

The local function for rule 101 is following:

\[
\varphi_{R101} = \left\{
	\begin{array}{lcl}
		1 & \mbox{if} & 110, 101, 010, 000 \\
		0 & \mbox{if} & 111, 100, 011, 001
	\end{array} \right. .
\]

In this case, $\varphi_{R101}$ has the same probability as $\varphi_{R86}$ to produce states one and zero. However $\varphi_{R101}$ is not a surjective rule and therefore has the Garden of Eden configurations, i.e., not all configurations have ancestors.

To classify global behaviour properly of $\varphi_{R86}$ and $\varphi_{R101}$ we should calculate their mean field polynomials. Mean field polynomial for $\varphi_{R86}$ is:

\begin{equation}
p_{t+1}=3p_{t}q_{t}^{2}+p_{t}^{2}q_{t}
\label{eqR86}
\end{equation}

\noindent and for $\varphi_{R101}$ we have:

\begin{equation}
p_{t+1}=2p_{t}^{2}q_{t}+p_{t}q_{t}^{2}+q_{t}^{3}.
\label{eqR101}
\end{equation}

\begin{figure}[th]
\begin{center}
\subfigure[]{\scalebox{0.33}{\includegraphics{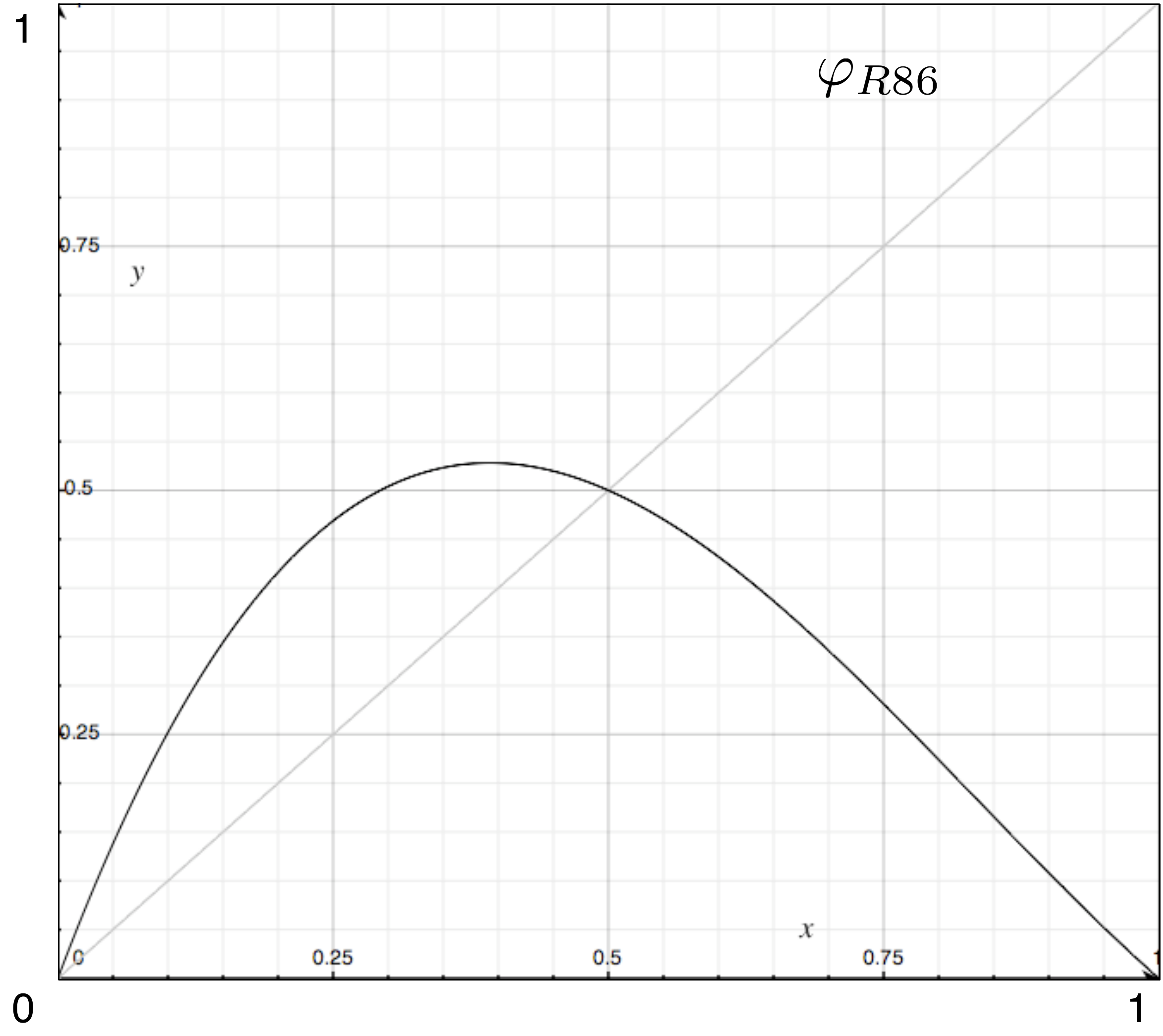}}} \hspace{1cm}
\subfigure[]{\scalebox{0.33}{\includegraphics{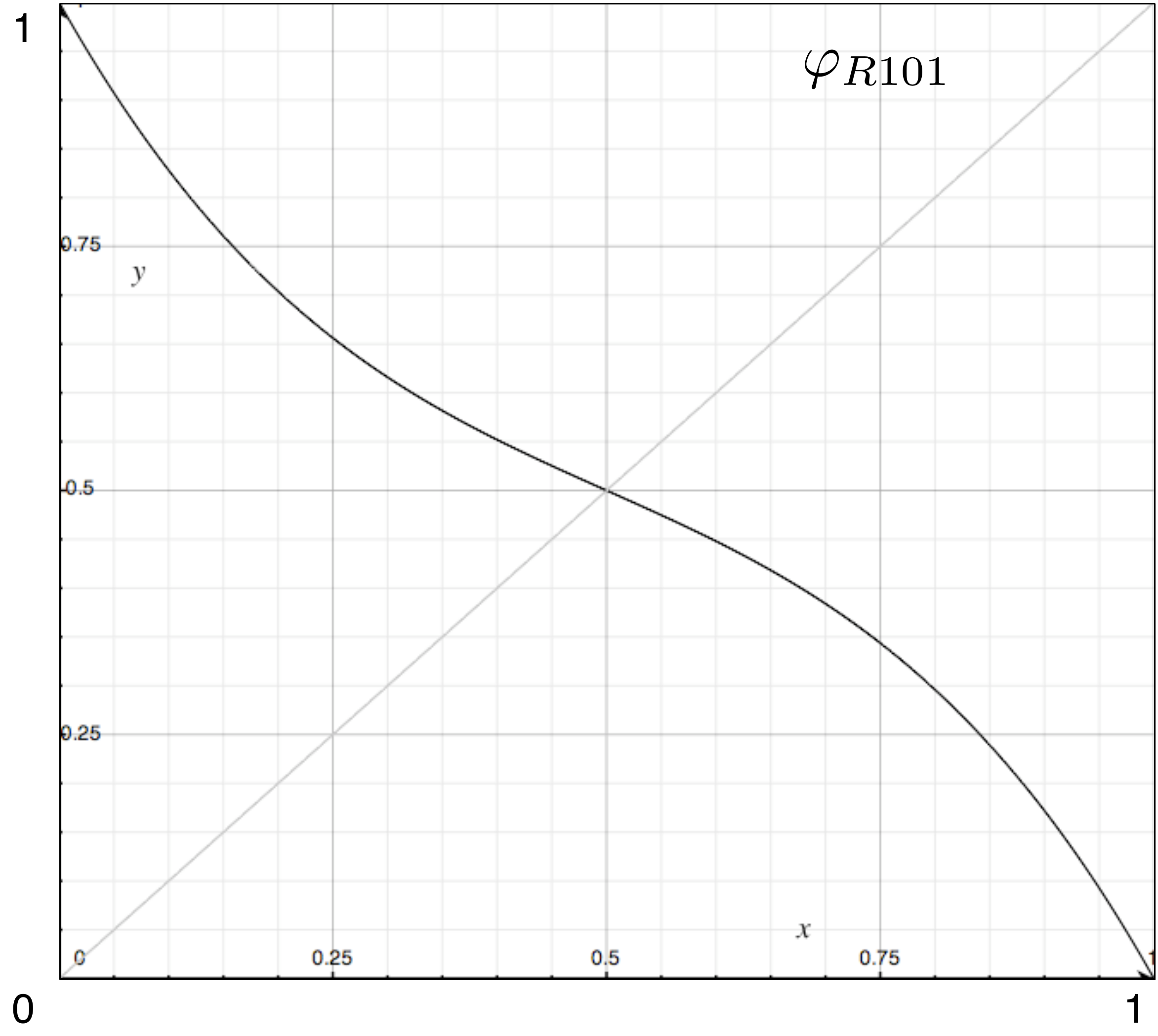}}}
\end{center}
\caption{Mean field curves for (a) $\varphi_{R86}$ and (b) $\varphi_{R101}$ respectively.}
\label{meanField}
\end{figure}

The polynomial for $\varphi_{R86}$ satisfies the mean field classification (Sect.~3). Where rules in CA class III do not have tangencies and therefore the curve crosses the identity. Consequently, $\varphi_{R86}$ evolves with a chaotic global behaviour (see Fig.~\ref{meanField}(a)).

This mean field polynomial has an stable fixed point when Eq.~\ref{eqR86} is $f=0.5$. This value relate the existence of densities where the population of cells in state one is preserved with few changes. Also such fixed point confirm its initial probability since $\varphi_{R86}$. Of course, if there are extreme densities of zeros and ones then next time the configuration will be filled of states zeros only, a homogeneous global state.

Mean field curve for $\varphi_{R101}$ (see Fig.~\ref{meanField}(b)) presents another characteristic. Again the curve does not cross the identity and its global behaviour $\Phi$ should then be chaotic. Its stable fixed point $f=0.5$ relates to the initial probability estimated since $\varphi_{R101}$. The curve displays what would happen if some initial configuration $c_0$ is dominated by state one, at the next step will be dominated by states zero and therefore this behaviour should repeat periodically. Such phenomenon also is balanced with its 50\% of density to each step.

Finally Fig.~\ref{chaosECA} displays two evolutions with typical chaotic behaviour in ECA. First evolution (a) displays the chaotic global evolution of $\varphi_{R86}$ since a random initial condition with a 50\% of density. That confirm an evolution without some order or pattern defined. Second evolution (b) displays the chaotic global behaviour for $\varphi_{R101}$ with the same parameters.

Now we will select a kind of memory and uncover ``hidden'' properties of chaotic ECA's.

\subsection{Filtering evolutions}
Filters selected in CA are a useful tool for understand ``hidden'' properties of CA. This tool was amply developed by Wuensche in a context of automatic classification of CA. The filters were derived from mechanical computation techniques \cite{HC97} and analysis of cell-state frequencies \cite{Wue99}.

Others derivations deducing filters relate as tiling, were reported for ECA rule 110 \cite{MMS06}, and rule 54 \cite{MAM06}. However in general such filters are not widely exploited in CA studies. We consider the tile representation to identify filters as block of cells in one or two dimensions. We  explain each tile filtering $\varphi_{R86}$ and $\varphi_{R101}$ in the Sec. 4.3.

\subsection{Complex dynamics emerging from $\varphi_{R86}$ and $\varphi_{R101}$ with majority memory}

\begin{figure*}
\centering
\includegraphics[width=6.5in]{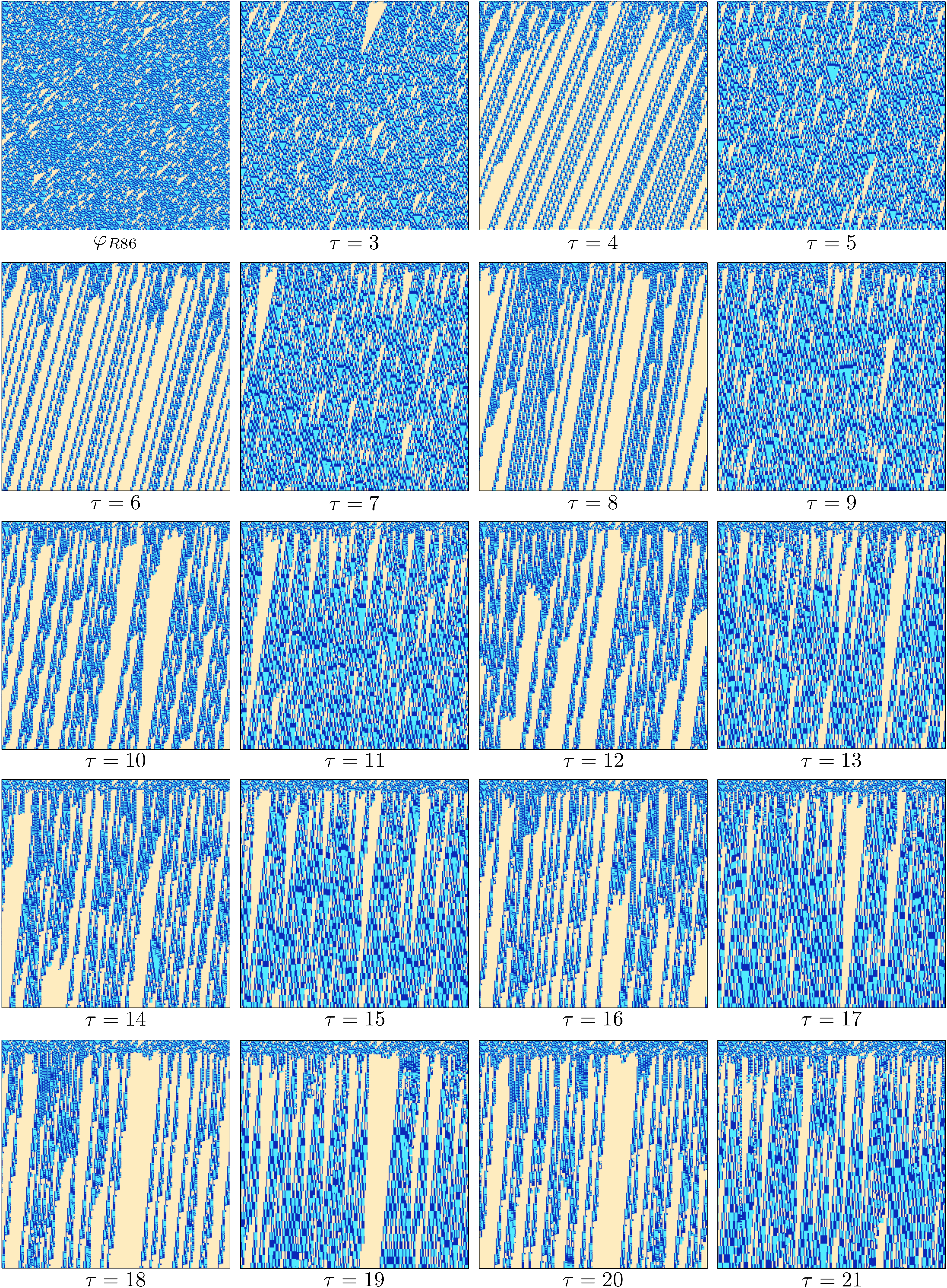}
\caption{Majority memory $\phi_{R86maj:\tau}$ working in $\varphi_{R86}$ with $\tau$ values of 3 to 21, respectively. The first one is the original ECA rule 86 evolution. All snapshots evolve with the same random initial condition of 50\% over an array of 300 cells to 300 generations, all evolutions are filtered.}
\label{majMem86}
\end{figure*}

\begin{figure*}
\centering
\includegraphics[width=6.5in]{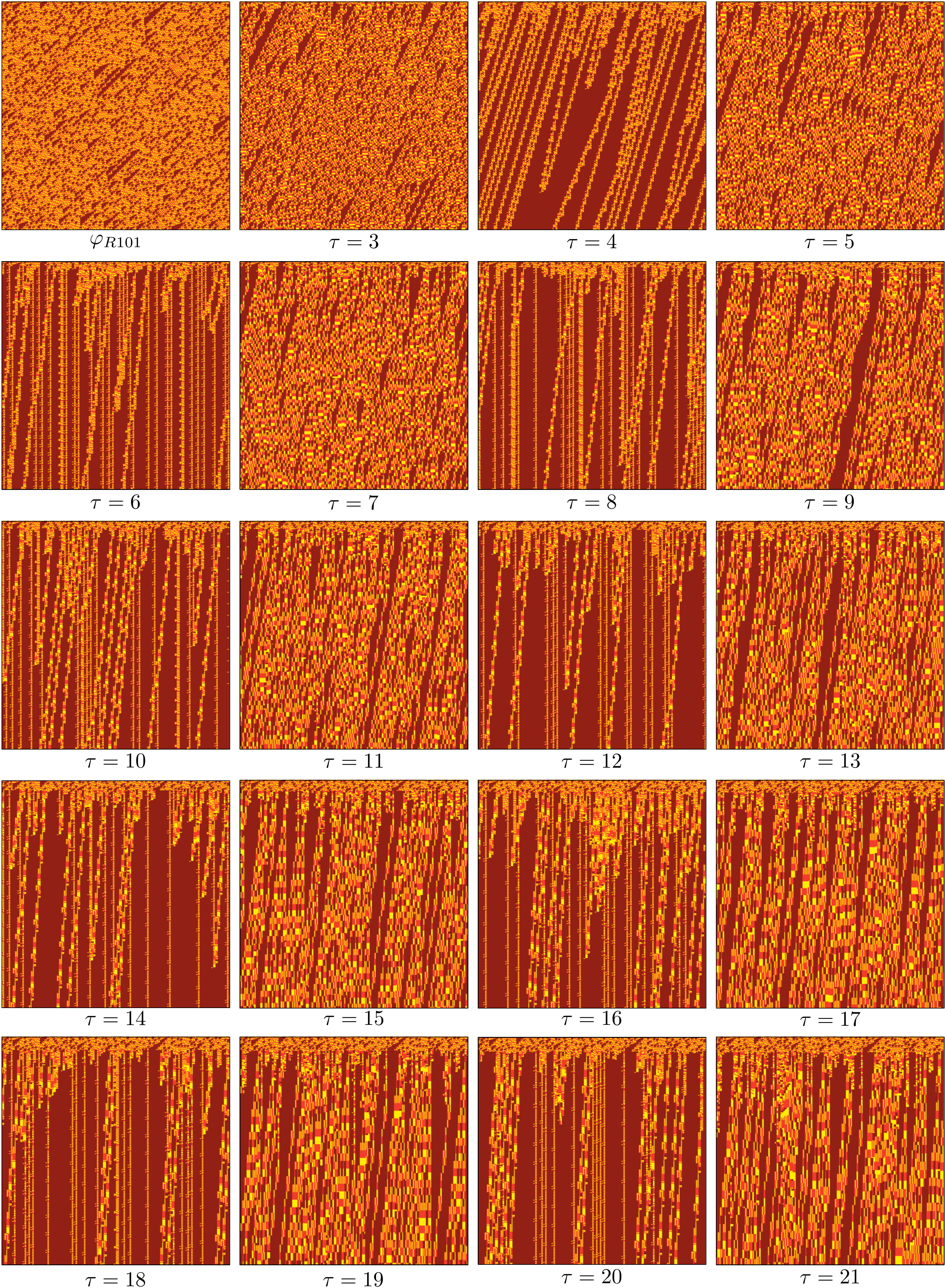}
\caption{Majority memory $\phi_{R101maj:\tau}$ working in $\varphi_{R101}$ with $\tau$ values of 3 to 21, respectively. The first one is the original ECA rule 101 evolution. All snapshots evolve with the same random initial condition to 50\% over an array of 300 cells to 300 generations, all evolutions are filtered.}
\label{majMem101}
\end{figure*}

Firstly we should consider a kind of memory, in this case the majority memory $\phi_{maj}$ (see Eq.~\ref{eq-majmem}) and then a value for $\tau$. This value represents the number of cells backward to consider in the memory (as we saw in Sec. 2.2).

Implementing the majority memory $\phi_{maj}$ we can select some ECA and experimentally explore its effect. Figures~\ref{majMem86} and~\ref{majMem101} show outcomes of selecting memory $\tau$ working on $\varphi_{R86}$ and $\varphi_{R101}$ respectively. The result is a new family of ECA but now with majority memory, they are the rules: $\phi_{R86maj:3}, \ldots, \phi_{R86maj:\infty}$, and $\phi_{R101maj:3}, \ldots, \phi_{R101maj:\infty}$.

As a characteristic while the memory is working on $\phi_{R86maj}$ and $\phi_{R101maj}$ a periodic background was more evident and it can be represented as a tile. These filters work as well on the original rules $\varphi$, see Fig.~\ref{majMem86} and~\ref{majMem101}.

The memory effect produces an emergency of patterns. The patterns interact quickly (in time-scale of CA development) with each other. In fact, for $\phi_{R86maj}$ and some values of $\tau$ the original behaviour changes its dynamics dramatically. Following
our previous findings \cite{MAA10}, in press] we consider only even values that offer better global dynamics. Thus the new rule $\phi_{R86maj:8}$ displays particles travelling in different velocities on a periodic background (see Fig.~\ref{majMem86}).

The second case $\phi_{R101maj}$ displays more attractive result. These three rules support stationary and mobile particles, travelling and colliding, some collisions can be interpreted as solitonic reaction \cite{Ada02} (see Fig.~\ref{majMem101}).

Also on all evolutions a filter was selected to clarify evolutions and patterns.\footnote{All evolutions simulated to ECA and ECA with memory they are calculated with OSXLCAU21 system, available from http://uncomp.uwe.ac.uk/genaro/OSXCASystems.html} Filters really are useful to recognize periodic dominant patterns of objects moving into such local universes.

The first two-dimensional tile working in $\phi_{R86maj}$ is represented as {\footnotesize $t_{\phi_{86}} = \begin{bmatrix} 101 \\ 101 \end{bmatrix}$}. Also this tile works on the original evolution rule as shows the Fig.~\ref{majMem86}. Tile reported for $\phi_{R101maj}$ is determined for the two-dimensional tile {\footnotesize $t_{\phi_{101}} = \begin{bmatrix} 100 \\ 100 \end{bmatrix}$}. So this filter works on the original evolution rule as well as shows the Fig.~\ref{majMem101}.

The effect of memory producing new evolution rules is preserved in some way. Initially the existence of a filter that can evolve on all different function, that is not rare because the memory only read the history and process the new generation with the original rule.

\subsection{Coding particles}

\subsubsection{Self-organization by structure formation}
Patterns as particles and non-trivial behaviour emerging in these new CA with memory $\phi_{R86maj}$ and $\phi_{R101maj}$, naturally conduce to well-know problems as self-organization.

\begin{figure}[th]
\centering
\includegraphics[width=3.5in]{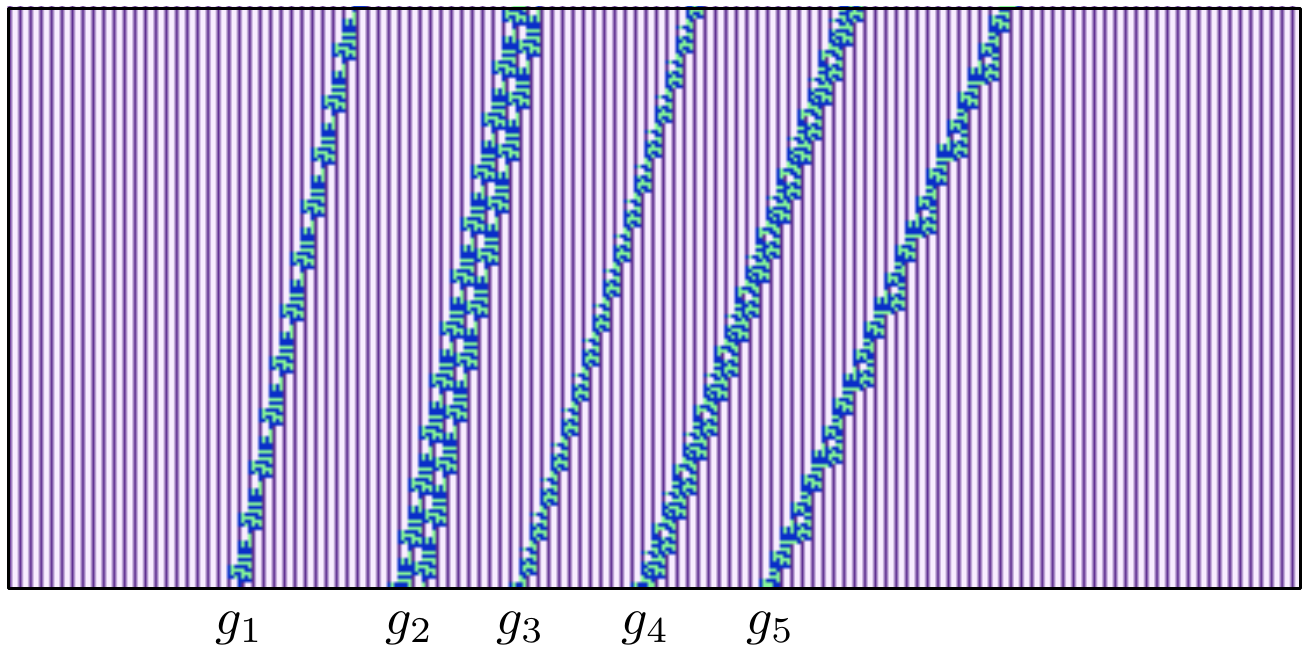}
\caption{Set of particles $\cal G$ emerging and living in $\phi_{R101maj:4}$.}
\label{particles}
\end{figure}

Considering the evolution rule $\phi_{R101maj:4}$, we have done a classification of particles in this local universe (see Fig.~\ref{particles}). The universe is not bigger compared with other complex rules. However that all particles in $\phi_{R101maj:4}$ can be produced from other particles in binary collisions. Such self-organization by structure formation \cite{Kau93} is demonstrated in the following set of reactions between particles:

$$
g_i \rightarrow g_j = g_k
$$

\noindent such that $i \neq j \neq k$ and $i,j,k \in \cal G$$_{\phi_{R101maj:4}}$. 

\begin{figure}[th]
\centering
\includegraphics[width=3.5in]{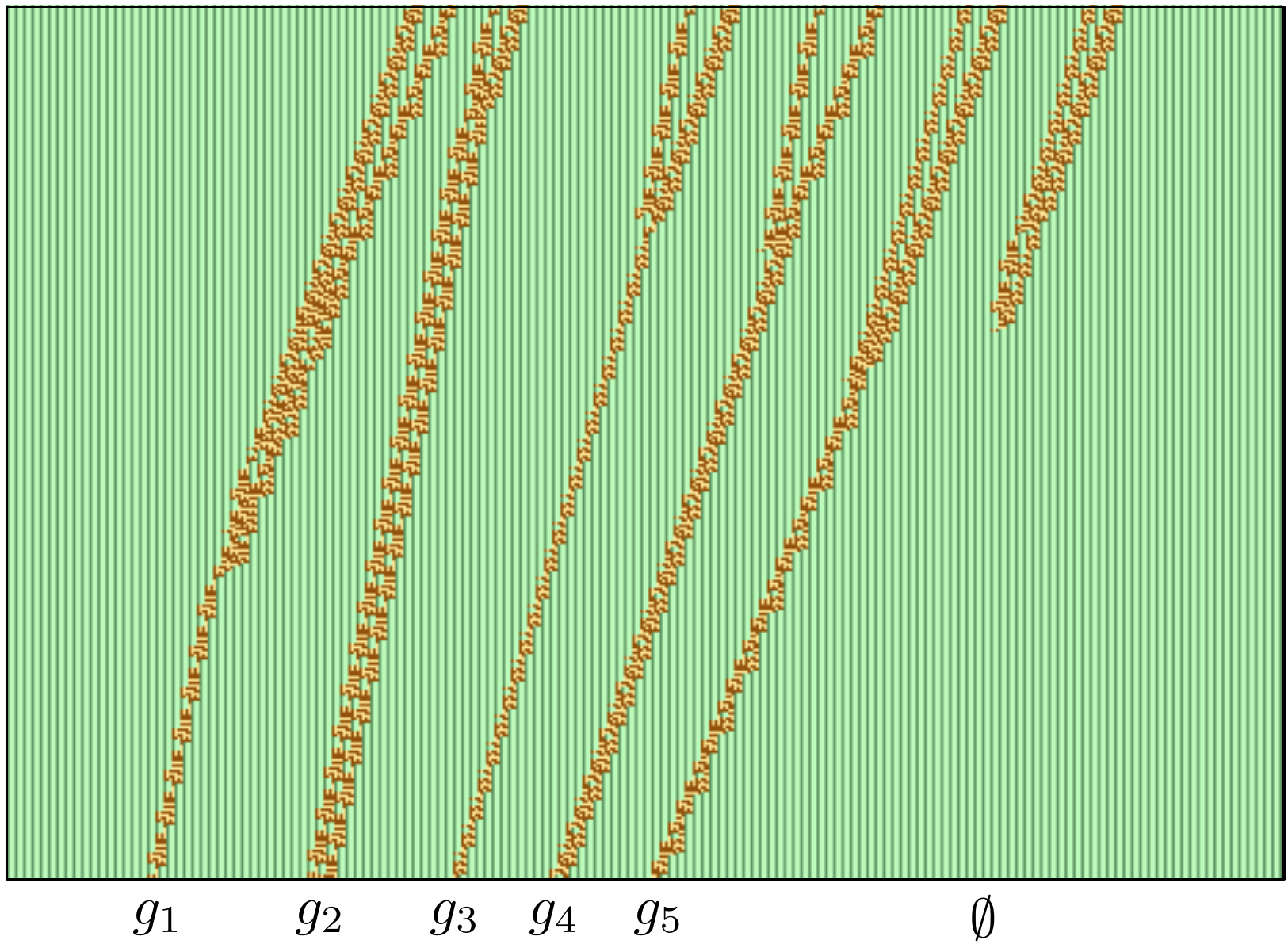}
\caption{Self-organization by particle collisions. The set $\cal G$$_{\phi_{R101maj:4}}$ is produced. CA evolution is filtered.}
\label{selfOrganizationR101}
\end{figure}

Figure~\ref{selfOrganizationR101} presents the set of reactions necessary to produce every particle:

\begin{quote}
\begin{enumerate}
\item[a)] $g_4-b^3-g_5 = g_1$
\item[b)] $g_1-b^2-g_4 = g_2$
\item[c)] $g_1-b^4-g_4 = g_3$
\item[d)] $g_1-b^6-g_5 = g_4$
\item[e)] $g_3-b^3-g_4 = g_5$
\item[f)] $g_3-b^2-g_4 = \emptyset$
\end{enumerate}
\end{quote}

Of course, they are not all possibilities to get every particle and a organization of several particles could be produce even more complex behaviour imitating physical, biological, chemical or computational phenomena: wave propagation, reaction-diffusion, morphogenesis, particle collision, fluid-dynamics, (tissue) grown, pattern formation, self-reproduction, self-assembly, artificial life, synthetic constructions (engineering), tessellation, differential equations, soliton solutions, formal languages, or unconventional computing \cite{Ada02}, \cite{Bar97}, \cite{Mar07}, \cite{Mit09}, and \cite{Mor98}.

\subsubsection{Generator pattern}

\begin{figure}[th]
\centering
\includegraphics[width=2.8in]{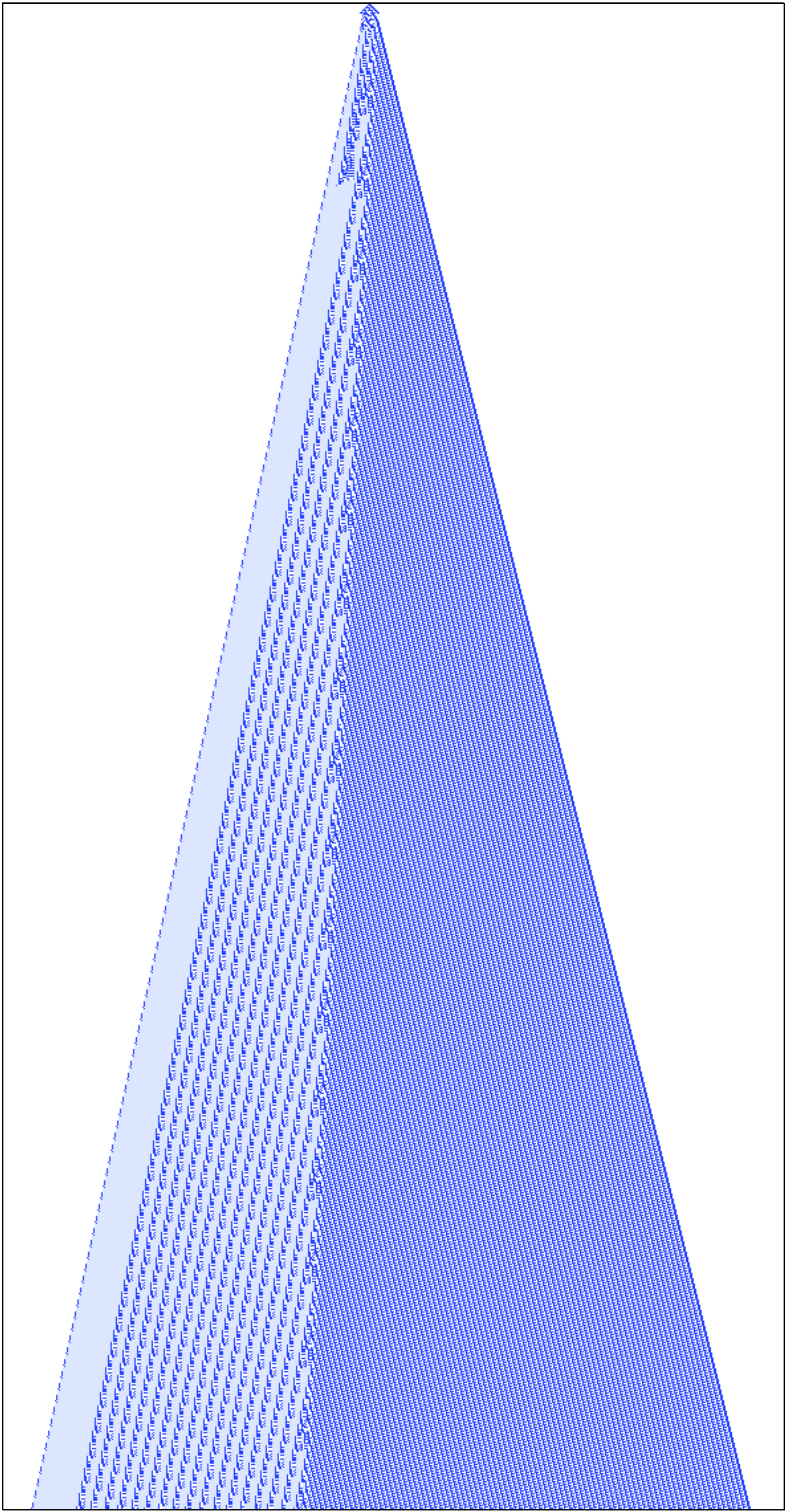}
\caption{Stream of particles and fuse patterns emerging from a single cell in state 1 with $\phi_{R86maj:8}$. These patterns exhibit unlimited growth.}
\label{streamParticlesR86}
\end{figure}

\begin{figure*}
\centering
\includegraphics[width=6.6in]{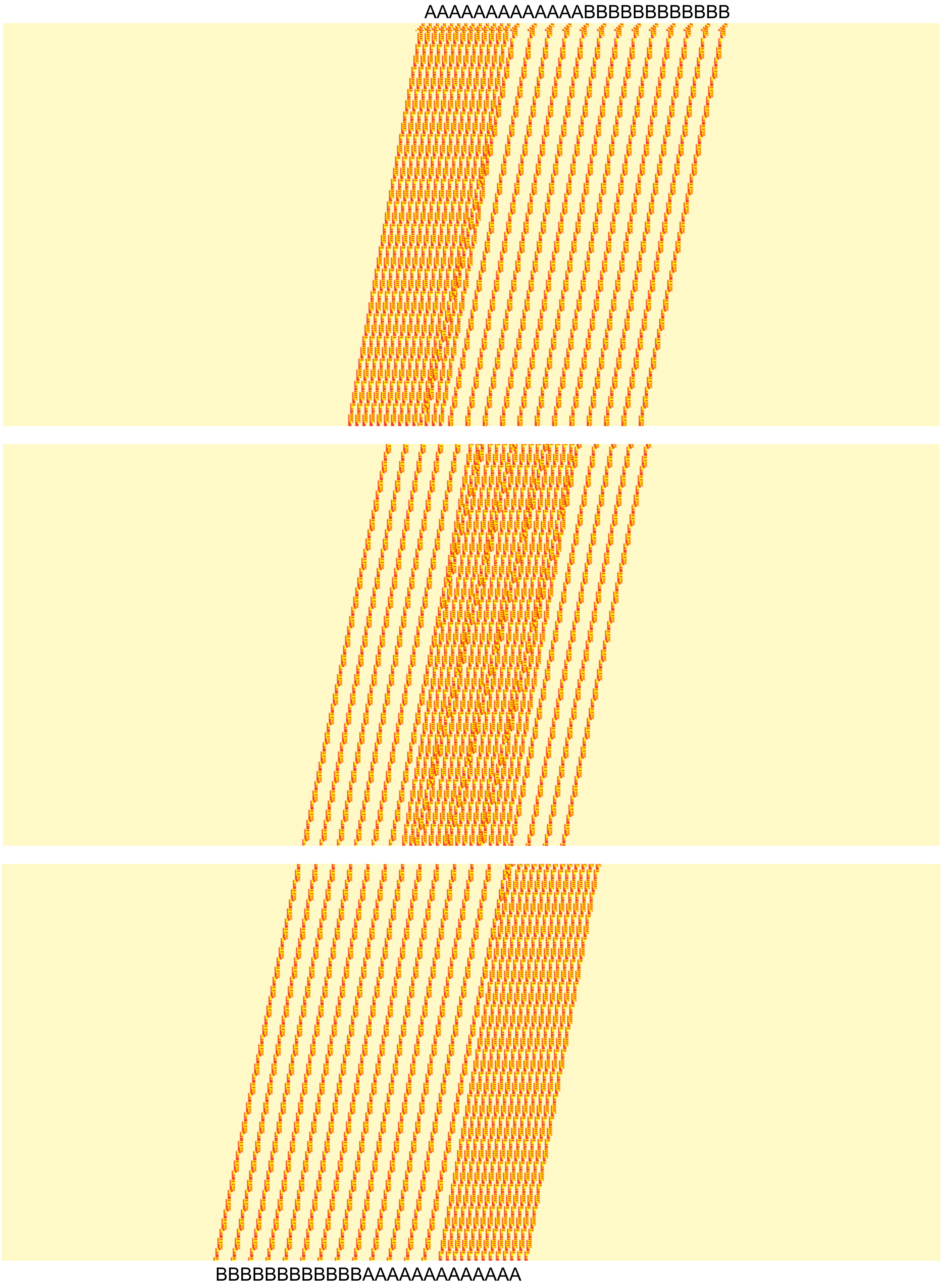}
\caption{A simple substitution system processing the word $A^{12}B^{12}$ to $B^{12}A^{12}$ with $\phi_{R86maj:8}$. The final production is reached on 6,888 generations by synchronization of multiple soliton reactions.}
\label{substitution}
\end{figure*}

Figure~\ref{streamParticlesR86} shows the evolution of CA from a single cell in state 1, the automaton is governed by rule $\phi_{R86maj:8}$. A fuse pattern is organized by stream of gliders (left) emitted periodically every 62 steps and a fixed periodic pattern (right) growing with a velocity of $-\frac{1}{4}$.

The above examples of CA with memory are just two simple cases showing the memory effect on traditional chaotic functions. Another cases were developed for the ECA rules 30 and 126 in \cite{MAS10}.

\subsubsection{Implementing basic computing functions}
\begin{figure}[th]
\centering
\includegraphics[width=2in]{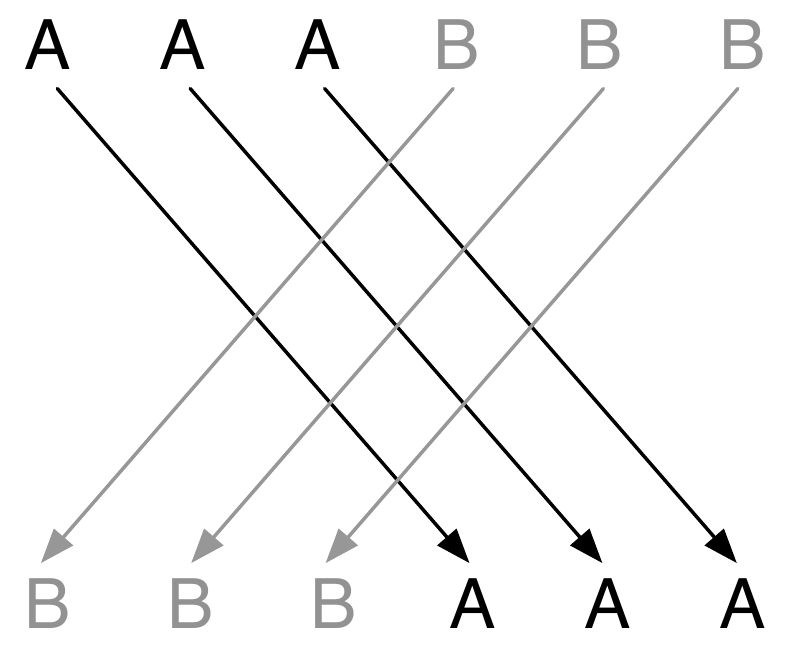}
\caption{Schematic diagram adding the string $w_2$ to head of the list $w_3$.}
\label{substitutionDiagram}
\end{figure}

We can employ the particles codification to represent solutions of some basic computing functions. Let us consider the rule $\phi_{R86maj:8}$. We want to implement a simple substitution function {\sf addToHead} working on two strings $w_1=A_1,\ldots,A_n$ and $w_2=B_1,\ldots,B_m$, where $n,m \geq 1$. For example, if $w_1=AAA$, $w_2=BBB$ and $w_3=w_1w_2$ then the {\sf addToHead($|w_2|$)} will yield: $w_3=w_2w_1$ (see schematic diagram of Fig.~\ref{substitutionDiagram}).

To implement such function in $\phi_{R86maj:8}$ we must represent every data `quantum' as a particle. Gliders $g_1$ and $g_2$  are coded to reproduce a soliton reaction.\footnote{These gliders are a reflection of $\phi_{R30maj:8}$, because ECA rule 86 is the reflection of rule 30, and consequently their gliders emerging with memory can be coded in a similar way \cite{MAA10}.} Another problem is synchronize several gliders and obtain the same result with multiple collisions.

The codification is not sophisticated however a systematic analysis of reactions is required. We known than a periodic gap and one fixed phase between particles is sufficient to reproduce the {\sf addToHead} function for any string $A^nB^m$.

Figure~\ref{substitution} shows fragments of evolutions of $\phi_{R86maj:8}$ from an initial condition coded by gliders, representing the string $AAAAAAAAAAAABBBBBBBBBBBB$. Using function {\sf addToHead} we produce the final string $BBBBBBBBBBBBAAAAAAAAAAAA$ after 6,888 generations. The first snapshot in Fig.~\ref{substitution} shows its initial configuration and the first 400 steps, the middle snapshot mainly presents how the string $w_1$ across the string $w_2$ preserving the information (soliton reaction), and the third snapshot shows the final global configuration so given the string $w_2w_1$ processed in parallel with $\phi_{R86maj:8}$.

\section{Discussion}
We have demonstrated that elementary cellular automata (ECA) with memory offer a powerful approach to discovering complex dynamics based on particles and non-trivial reactions between the particles. Such problem has been substantiated by a number of different techniques, e.g. number-conservation \cite{BF02}, \cite{III04}, exhaustive search \cite{Epp02}, tiling \cite{MMS06}, \cite{Mar07}, de Bruijn diagrams \cite{MAM08}, $Z$-parameter \cite{Wue99}, genetic algorithms \cite{DMC94}, mean field theory \cite{Mc90} or from a differential equations point view \cite{Chu07}. Thus the memory function $\phi$ offers a more easy way to get similar and, in some cases, more strong results reporting new complex rules in ECA with memory.

\begin{figure}[th]
\centering
\includegraphics[width=3.2in]{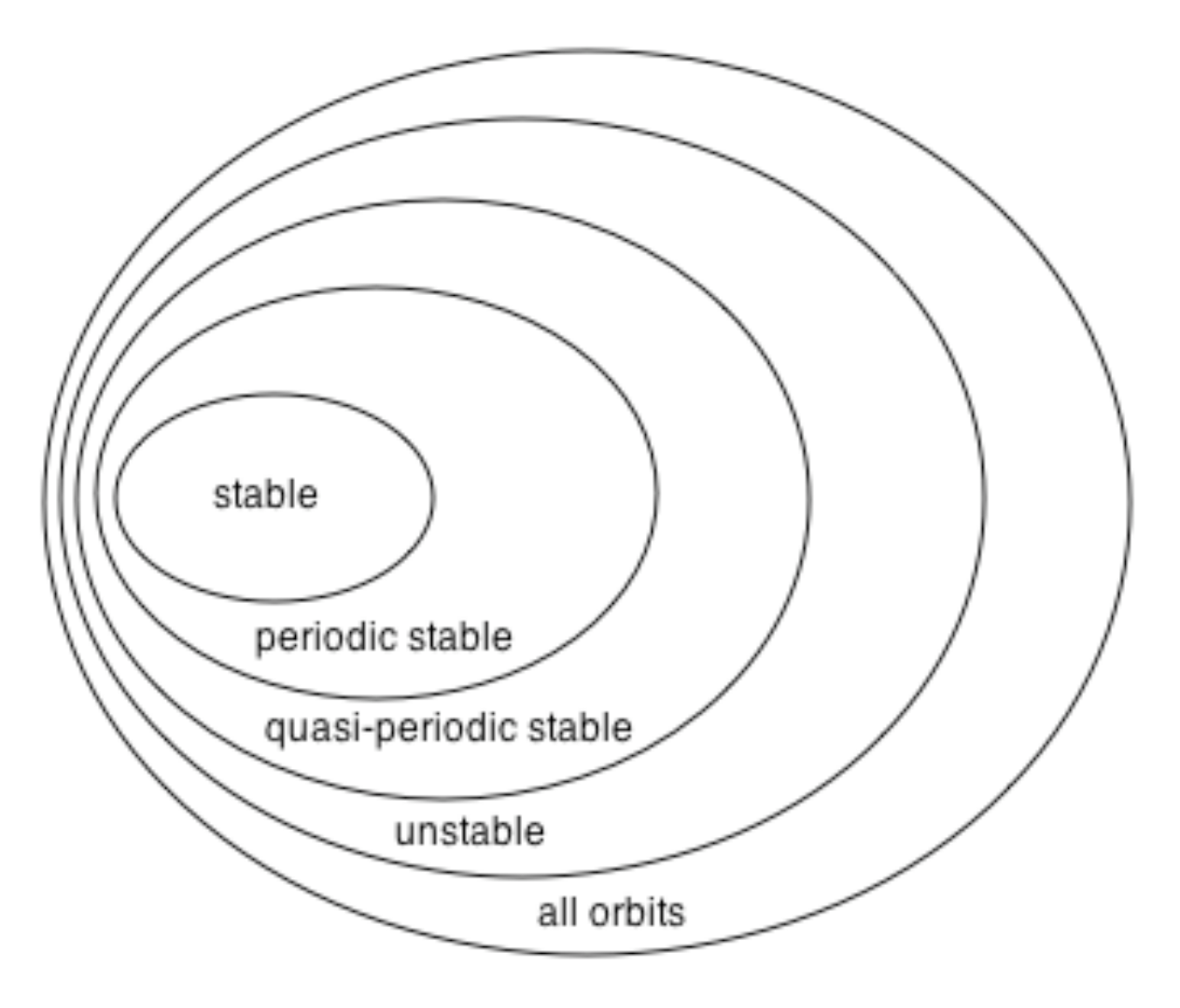}
\caption{Classes of global behaviour.}
\label{classes}
\end{figure}

We have enriched some classic chaotic ECA rules with majority memory and demonstrated that by applying certain filtering procedures we can extract rich dynamics of travelling localizations. Therefore, we can deduce a relation on chaotic systems decomposed in complex dynamics as a self-contained set. Generally a relation of sets of complex dynamics can be self-contained describing $\Phi$ as attractors, like a set diagram (Fig.~\ref{classes} \cite{AMS06}).

This way, the most bigger set in Fig.~\ref{classes} `all orbits' corresponds to complex dynamics and the `unstable' set represents the chaotic systems. Indeed there is a number of properties between orbits and characteristics that cannot be inferred directly. However the memory plays a role of a powerful tool to discover such properties. Finally, the memory function $\phi$ can be applied to any CA or dynamical system.

\nonumsection{Acknowledgments}
\noindent G. J. Mart\'{\i}nez is supported by EPSRC grant EP/F054343/1 and R. Alonso-Sanz by EPSRC grant EP/E049281/1.


\end{document}